# Dissipating with Relations: Implication for the Entity-Relationship Model

Sabah Al-Fedaghi
line Computer Engineering Department
Kuwait University
Kuwait
sabah.alfedaghi@ku.edu.kw

*Abstract*—Difficulties arise when conceptual modeling lacks ontological clarity and rigorous definitions, which is especially the case in the relationship construct. Evidence shows that use of relationships is often problematic when it comes to communicating the form of meaning of an application domain. Research on this topic is important because relationships are central to a number of approaches and commonly used by practitioners. In this paper, we study the notion of relation or relationship in the context of conceptual modeling. Specifically, we focus on the notion of relationship used in the entity-relationship (ER) model. The ER model is scrutinized through a new form of conceptual modeling called the thinging machine (TM) to pursue further understanding of the semantics of the relationship concept. The ER model is composed of three fundamental categories (i.e., entity, relationship and attribute), whereas TM is built from one ontological category called the thing/machine (thimac). Several ER diagrams are re-casted as TM diagrams, creating a categorical collision with interesting implications regarding the status of the conception of relationship in a conceptual model. The re-modeling shows that the relational construct is dissipated into TM flows of things and chronology of events.

## I. INTRODUCTION

A problem-solving process begins with recognition of a problem and ends with implementation of a solution, and one of the tools used within such a process is modeling. Modeling as part of a broader process is critical since it facilitates communication "with nonmodelers about the assumptions that go into the model and the intuition behind the model's results" [1]. In software engineering, models help designers think about problems and solutions within the technical team or with all the stakeholders [2]. In this context, "conceptual modeling is the activity of describing problem structures in a way that still is relatively independent from the technology and strategy used to solve the problem" [3]. Such a description involves representing aspects of the physical and abstract world to understand the depicted reality [4]. Modeling is also used in documentation, analysis, and design of the modeled system. Conceptual models offer a contrast to models used in science and engineering because "they don't model the world, but rather our conceptualizations of the world" [5]. For example, "engineers use a three-dimensional model, but there is no representation of the concepts 'wall', window', etc. On the other hand, we may build a conceptual model for a building by asking occupants to describe their conceptualizations of the building" [5].

Difficulties arise regarding conceptual modeling that "lacks ontological clarity when a one-one mapping does not exist between conceptual modeling constructs and real-world constructs" [6] (e.g., one real-world construct [an event] to be represented by more than one construct [an entity and a relationship]) [6]. Most conceptual modeling approaches are concerned with entities and relationships among entities. More recently, these two constructs have also featured in the object-oriented (OO) approaches:

> The entity construct is often replaced by the object construct, although both constructs have much in common. The traditional notion of a relationship, however, is used in only some OO approaches. It has much in common with the message-passing construct that is ubiquitous in OO modeling. [7]

### A. Problem: Relations

According to Wand, Storey and Weber [7],

> Often, rigorous definitions of [conceptual modeling] constructs are missing. This situation occurs especially in the case of the relationship construct. Empirical evidence shows that use of relationships is often problematical as a way of communicating the meaning of an application domain. For example, users of conceptual modeling methodologies are frequently confused about whether to show an association between things via a relationship, an entity, or an attribute.

In this paper, we study the construct of a relationship (relation), since it is claimed that relations are fundamental for conceptual modeling [8]. Guarino and Guizzardi [9-10] call the notion of relationship "one of conceptual modeling's most fundamental constructs." Research on the relationship construct is important because "it is central to many approaches and it is used extensively in practice" [6]. According to Burton-Jones and Weber [6], "the relationship construct is the linchpin through which humans associate different things in the world, thereby enabling them to conceptualize the subsystems and systems that are fundamental to their being able to reason about the world." In philosophy, from the perspective of a relational ontology, relations between entities are ontologically more fundamental than the entities themselves [11].

### B. Focus: The Entity-Relationship Model

Specifically, we focus on the construct of relationship as used in entity-relationship (ER) modeling. Chen [12] introduced the ER model, which can achieve a high degree of data independence and is based on "set theory and relation theory." The ER model has since been enhanced with additional constructs to form the extended ER [13].

ER diagrams have become popular, and they are central to most system-development methodologies, with numerous texts introducing ER modeling for the use of systems analysts and computer scientists [14]. According to [14], the "ER modelling is very weak – but it is very much easier to learn than the predicate calculus."

In ER modeling, an entity is a thing that can be distinctly identified (e.g., a specific person), and a relationship is an association among entities (e.g., a father and son). Chen [12] describes relationships in ER diagrams in the following manner:

> It is possible that some people may view something (e.g., marriage) as an entity while other people may view it as a relationship. We think that this is a decision which should be made by the Enterprise Administrator.

Modeling the semantics of organizations has made use of various versions of ER diagrams. According to [15], "The results of this approach have been decidedly mixed. The kind of data models created by database designers are concerned more with database artifacts (tables, keys, etc.) than with the underlying semantics of the business being modeled."

In this paper, we examine the notion of relation by inspecting the ER model. Furthermore, we explore the ER model through a new conceptual modeling called the thinging machine (TM). TM and ER modeling have different categorical systems. A system of categories provides an inventory of what constitutes the portion of reality represented by the conceptual model. ER modeling is composed of three types of fundamental categories: entity, relationship, and attribute. TM modeling has one ontological category called the thimac. It is sufficient to state now that the thimac is an entity that unifies the notions of object and process in one category. Section 3 will review the thimac concept and the TM model in general.

Exploring ER modeling through the TM model involves modeling of modeling, since TM modeling will be derived from existing ER diagrams as way to understand the semantics of the ER model. Accordingly, several ER diagrams are remodeled as TM diagrams. Since the literature about relations and relationships in modeling and philosophy is highly extensive, in the next section, we will present a glimpse of the notion of relation. Philosophy is needed here to understand the history of the concept of relation. Section 3 contains an enhanced review of the TM model, but the example in the section is a new contribution. The remaining sections present examples of remodeling of ER diagrams into TM notation.

## II. GLIMPSE REVIEW OF RELATIONS

Relations are an everyday part of life that connect people and things in the world around us. The notion of relation can be examined briefly from two perspectives: the relation construct in modeling and the concept of relation at large.

### A. Relations in Modeling

Chen [12] claims that "there are aspects of the intrinsic nature of 'real-world' that would justify distinction between entities and relationship." In addition, he claims that "nouns converted from a verb" (i.e., verb nominalizations) correspond to relationships (e.g., the shipping of the product to a customer or the assigning of an employee to a machine). According to Guarino and Guizzardi [9], relationships are objects (endurants). Moreover, relationships can bear properties (e.g., a project–worker relationship can have the attribute percentage of time as an intrinsic property of the relationship itself).

### B. Relations at Large

Philosophers have tried to understand relations theoretically and systematically, however, "philosophical theories of relations have proved extraordinarily difficult to construct ... and a unified philosophical account seems out of the question" [11]. According to Wildman [11], "We need to know whether relations are ontologically real or merely attributions made by conscious entities and expressed in language."

For Aristotle (the *Categories*), there were nine categories: quantity, quality, relation, habitus, time, location, situation (or position), action, and passion. Relations comprise one of the accidental categories; thus, relations are ontologically subordinate to entities. Aristotle prefers to speak of relations "as inhering in one thing and somehow pointing toward another" [16] (e.g., "taller" is a relative term because when we assert something is taller, we necessarily do so in comparison to something else [16]). Aristotle rejected this definition on the grounds that it allows certain substances (e.g., $x$ is ahead of $y$) to qualify as relations [16]. He requires relations to serve to *relate* two (or more) things not *stands* in some relation as in $x$ is ahead of $y$. Thus, there is a difference between relative terms such as "father" and "son" and others such as "mover" and "moved," "head" and "headed." Fatherhood is conceived of as if it were a kind of medium connecting a father with his son [16].

Avicenna (980-1037) argued that fatherhood is in the father alone and not in the son, and the same thing holds of sonship. Therefore, one must hold that there are two relations. Medieval philosophers recognize that predications involving relative terms (such as "$x$ is taller") are incomplete in a way that monadic or absolute predications (such as "$x$ is white") are not.

Currently, in some philosophical circles, the relations between entities are ontologically more fundamental than the entities themselves: "Unfortunately, there is persistent confusion in almost all literature about relational ontology because the key idea of relation remains unclear" [11].

III. THE THINGING MACHINE MODEL

This section will briefly review the TM model to establish it as a tool to explore relations. A more elaborate discussion of the TM model's philosophical foundation can be found in [17-23].

Evolution of research in modeling has led to ontologies where ontology is defined by its fundamental categories as the constituents from which everything else is constructed. The TM ontology is based on one category called thimacs. Several proposals of one-category ontology have arisen in philosophy (e.g., there exist only concrete particulars and there exist only properties) [24]. According to Paul [25], "One category ontologies are deeply appealing, because their ontological simplicity gives them an unmatched elegance and sparseness." Furthermore, Paul [25] proposes building "the world from one simple kind of relation: composition." Paul [25] also rejected "the need for a fundamental division between object and property, [or] any need for a primitive connecting relation of 'instantiation.'" The composition is definable from a primitive, proper parthood that fuses its fundamental constituents. In the traditional ontologies, the world is built from "propertied and related" space-time regions.

The TM thimac is a fundamental capsule of dual modes of a static thing (e.g., philosophically, *being*) that is unfolded from its dynamic machine (e.g., philosophically, *becoming*). The machine is constructed from a (non-geometrical) configuration of at most five generic operations (creation, processing, releasing, transferring, and receiving) connected by links (trajectories) for "visiting" things. The unfolding *creation* stage provides "existence" or "thereness" for a thing to be changed (processed) and transported (releasing, transferring and receiving).

The trajectories (called "flows") connect thimacs and (structurally related, e.g., whole part) subthimacs. These other thimacs may, in their turn, be subthimacs of yet other thimacs, and so on for higher levels of thimacs. Accordingly, a thimac is a chunk of flow and events in a larger thimac. Events are composed of thimacs having time subthimacs and their chronology is an ordering of thimacs.

A thimac can also be described as a categorical wrapper that embraces classical objects or processes exist simultaneously as two-fold entity. A thimac is an object (called a *thing*) and (in the broad sense) a process (called a *machine*); thus, the name "thimac." The thimac includes being-ness (e.g., a thing) and machine-ness (e.g., a process). Thus, the model is built from "thing/machines."

The thimac notion as a one-category ontology is not new (see [25]). In physics, subatomic entities must be regarded as particles and waves to describe and explain observed phenomena [26]. According to Sfard [27], abstract notions can be conceived in two fundamentally different ways: structurally, as objects/things (static constructs), and operationally, as processes. Thus, distinguishing between form and content and between process and object is popular, but "like waves and particles, they have to be united in order to appreciate light" [28]. The TM model adopts the notion of duality in conceptual modeling, generalizing it beyond mathematics.

In a thimac's two modes of being, "structural conception" means seeing a notion as an entity with a recognizable internal configuration and specified trajectories of flow. In the TM model, a thing can flow to another thing's machine. We consider such flow as a type of *relation* among thimacs. The behavioral way of conceiving thimacs emphasizes the dynamic aspects in terms of events (thimacs embrace time machines). Accordingly, we can identify a chronology of events to specify the accepted behavior. Thus, the events form another type of *relation* between thimacs. Therefore, as will be discussed later, the so-called relations are either flows or orders of events.

The term "thing" relies more on Heidegger's [29] notion of "things" than it does on the notion of objects. According to Heidegger [29], a thing is self-sustained, self-supporting, or independent—something that stands on its own. A thing "things"; that is, it gathers, unites, or ties together its constituents in the same way that a bridge unifies environmental aspects (e.g., a stream, its banks, and the surrounding landscape). The TM model goes beyond Heidegger's [29] being-ness to incorporate machine-ness.

The term "machine" refers to a special abstract machine called a "thinging machine" (see Fig. 1) that encapsulates the laws of flows. A TM is built under the postulation that only five generic actions/operations are performed on things: creating, processing (in the sense of changing), releasing, transferring, and receiving. A thing is defined as that which is created, processed, released, transferred, and/or received. A machine is defined as that which creates, processes, releases, transfers, and/or receives things. Since a thimac is a thing and a machine at the same time, we will alternate between the terms "thimac," "thing," and "machine" according to the context. If a certain thimac exists, then it is possible that it may be a subthimac of other thimacs, and its partial thimacs (which do not include all stages) are primitival thimacs (i.e., they do not include any subthimacs). An example of a partial thimac is a thimac with a single stage, say, create (∃ in logic). In this case the thimac does not include the process, release, transfer and the receive stages.

The five TM flow operations (also called stages) form the foundation for thimacs. Among the five stages, the flow (a solid arrow in Fig. 1) of a thing means the trajectory of a thing's "motion," which occupies different stages. The arrow represents a projected flow just as, say, the path of the Nile on a map. The TM diagram reflects the succession imposed on the "motion" of a thing: create→release→transfer, etc. The flow among the five stages is the law of flow though the thimac. The flow is the occupation of different stages at different times.

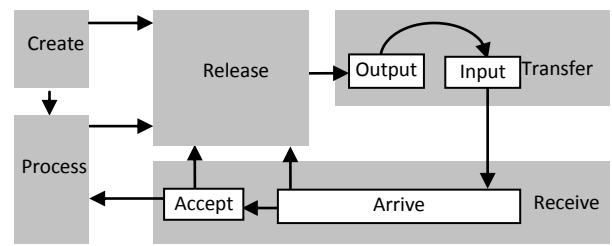

Fig. 1. A thinging machine.

In a TM model, a thing has no other place to be besides one of the five generic stages of the machine. Thus, thimacs are arenas through which things flow. The flow establishes relations among thimacs.

The generic TM flow operations can be described as follows:

- *Arrival*: A thing occupies the first stage (input gate) of a new machine.
- *Acceptance*: A thing is permitted to occupy the *accept* stage in the machine. If arriving things are always accepted, then arrival and acceptance can be combined to become the "receive" stage. For simplicity, this paper's examples assume a receive stage.
- *Release*: A thing occupies a release stage where it is marked as ready to be transferred outside of the machine.
- *Transference*: The transference is as the ground-beam of the door through which things flow thus encompassing both input and output. A thing occupies a transfer stage (in and out gate/port) to be transported somewhere outside/inside of the machine. It is Heidegger's "threshold" that bears the "between" where the inside and outside *penetrate* each other [30]. In Shannon's communication model, the *transfer* stages constitute double "doors" as source and destination connections on both ends of the communication channel. This is in analogy to two adjacent hotel rooms with two doors between them, each having a lock of its own door leading to the other room.
- *Creation*: A new thing is born (created) in a machine. A machine creates, in the sense that it finds or originates a thing; it brings a thing into the system and then becomes aware of it. "Creation" can designate "bringing into existence" in the system because what exists is what is found.

In addition, the TM model includes memory that is accessed from all stages and triggering (represented as dashed arrows) that connects thimacs in non-flow ways (e.g., classical control flow among independent programs that have no data flow among them).

Note that the TM model is by no means a new philosophical system (e.g., ontology for reality). Rather, it is a conceptual tool (e.g., as flowcharting) for modeling systems in software engineering (e.g., as object-oriented UML). Its conceptualization process touches reality because computerized systems—as with everything else—are rooted in reality. The TM model may borrow expressions and terms from philosophy, but it lacks philosophical scrutiny and formality. However, the TM model brings new ideas in software engineering that are already present in philosophy, such as the one-category ontology. The five generic TM operations seem to be a new contribution that still needs to be proven.

**Example** (From [31]): Consider the ER diagram shown in Fig. 2 that represents a marriage relationship. Fig. 3 shows the corresponding TM model. A person (Circle 1) flows (2) to husband (3) and wife (4) to be processed (5) to create marriage (6). A person, husband, wife, and marriage are thimacs. Inside the thimac, a person is presented as a *thing* while the whole box titled "person" (thick circumference) is a *machine*. Under such conceptualization, a person is a thing that flows and a machine that has the submachines ID and husband or wife.

At this point, we need to introduce time in the model through the notion of event. The event is a thimac (e.g., Fig. 4 shows the events of the example). The date of the marriage is a time thing that cannot be incorporated in the static description of Fig. 3. The event is a thimac and can be specified as follows.

Event 1 ($E_1$): There is a person (as in logic, ∃ a person).
Event 2 ($E_2$): A person moves to be a husband.
Event 3 ($E_3$): A person moves to be a wife.
Event 4 ($E_4$): The persons who move to be a husband and wife are processed.
Event 5 ($E_5$): Marriage is established.

Each subdiagram of Fig. 3 (the static model) inside an event is called the region of the event. The events may have other subthimacs besides time and region thimacs. Note that *date* is a time subthimac in the marriage thimac. Fig. 5 shows the chronology of events in the marriage relationship.

The TM representation seems to be richer (more detailed) than the ER model. In the remaining part of this paper, for simplification's sake, we will represent an event only by its region.

IV. RELATIONS OR EVENTS?

As mentioned in the introduction, according to Wand and Weber [32], relationships are often problematic as a way of communicating the meaning, e.g., "users are frequently confused about whether to show an association between things via a relationship, an entity, or an attribute." The following subsection presents an example of the involved difficulties.

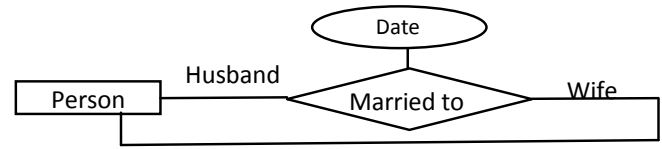

Fig. 2. The ER model of a marriage (Redrawn from [31]).

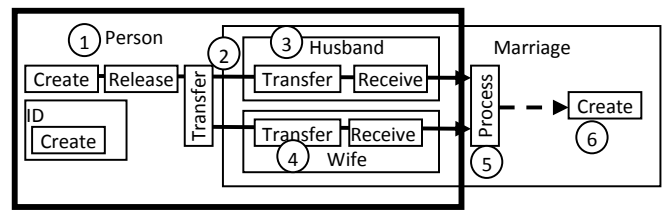

Fig. 3. The TM model of a marriage relationship.

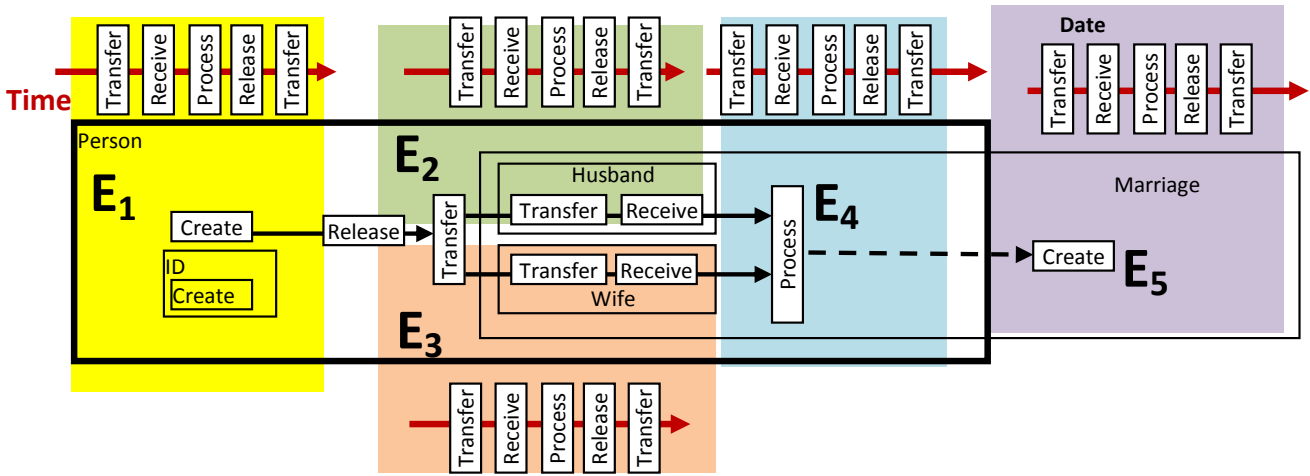

Fig. 4. The events in the TM model of a marriage relationship.

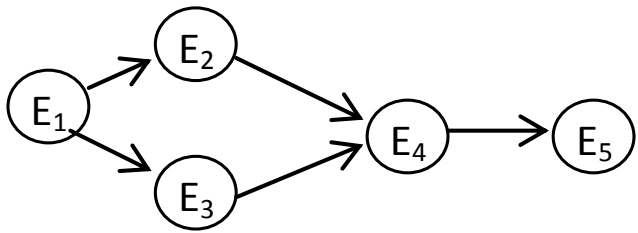

Fig. 5. The chronology of events in the TM model of a marriage relationship.

*A. ER Modeling and Time*

Wand and Weber [32] give an example of ER representation of A student attends a university, as shown in Fig. 6, which indicates that "things of class student possess a mutual property with things of class university–namely, an 'attends' mutual property." This implies that the university has the "attends" property. In addition, they explain that "the possible states of student things and the possible states of university things are not independent of each other." Hence, this implies "the existence of a mutual property." The state is the set of values at a certain point t and an event is a change of the state of a thing, but not a thing. The world is made of things that possesses properties, and properties are always attached to things [32].

*A student attends a university* can interpreted in two ways:

- A student becomes a member of *S* (student ε *S*), where *S* is the set of university students. Thus, the diamond shape "attends" in Fig. 6 can be replaced by "becomes a member of."
- Since we are constructing a *data* model, a student's *record* is added to the student database (set of records).

In both cases, it is trivial to indicate that the relationship between a student and *S* (record and file: a set of tuples) is one (student) to many (university students).

Figs. 7 and 8 show the TM model of two interpretations, respectively. In Fig. 7, the student (Circle 1) joins the set of students of the university (2). In Fig. 8, the student (1) attendance record is constructed (2). The record flows (3) to the university (4) database (5). In the TM model, the student and university are thimacs. The attendance record as a thing flows to the database (a machine), where it is stored. The TM model has a composition in terms of thimac/subthimac and flow that *relates* the student to the university. This is a static description that does not involve time.

The ER diagram seems to be a shorthand notation of the TM model. Fig. 9 shows how to produce the ER diagram by eliminating the flow of the attendance to the university system and keeping only the most common labels of *student* and *university*. Accordingly, starting with the TM diagram in Fig. 9, the top diagram in Fig. 8 (the ER diagram) is produced. Then the shape of the attendance is changed, producing the middle diagram. Lastly, the student is attached to the diamond shape to produce the bottom figure.

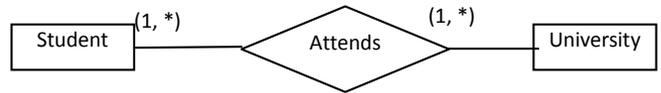

Fig. 6. The ER model of *A student attends a university* [32].

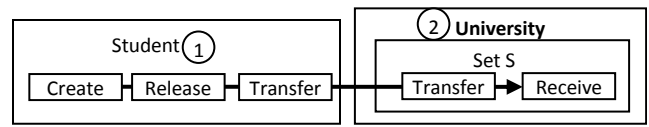

Fig. 7. The TM representation of *A student attends a university*.

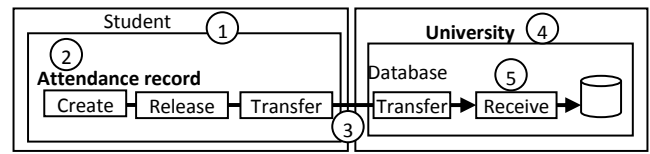

Fig. 8. The TM representation of *A student attends a university*.

Wand and Weber [32] then considered the statement *The student attends the university from a certain date*, which is modeled as Fig. 6, but the relationship designator *attends* has the start date as an attribute. They observed, "this representation is problematical because its interpretation is ambiguous. Specifically, start date is a mutual property of the student and university things." In the ER model, therefore, it should be shown using a relationship symbol if construct overload is to be avoided. Instead, it is shown in the ER diagram using the intrinsic attribute symbol. How then should the ER diagram be interpreted? According to Wand and Weber [32], Burton-Jones and Weber [6] developed the model in terms of two relations (i.e., start date and attends) and showed that such a model is "better able to undertake problem-solving in a domain" than an ER model with start date as an attribute of the relationship.

Alternatively, the dynamic TM model is shown in Fig. 10. In the dynamic model, the start date is an "attribute" (subthimac) of an event and not a relation. A certain value of time is associated with an event as a thimac that embodies a time subthimac. It seems that such an observation does not need more justification.

### B. Modeling Events

. For Wand and Weber [32], an event *is a change of state of thing*. Many scholars have suggested ways to incorporate events in the ER model. Modeling events as process relationships between entities has been proposed (e.g., "the completion of a batch of 20 purchase requisitions triggers an event that causes a purchase requisition to be approved or rejected") [33].

Chen [12] modeled the relationships to bear properties (e.g., *a project–worker relationship can have the attribute percentage-of-time* as an intrinsic property of the relationship itself). According to Guarino and Guizzardi [9], "Chen is correct in admitting that relationships can bear properties." This is not the case from the TM perspective.

Fig. 11 shows the TM model of *A worker who works on a project (in addition to his or her work)*. In the figure, the worker works in his or her regular duties and on a project. This is a static model that has nothing to do with time, quantity, or otherwise. Percentage of time, if it is a property or a relation, then appears when the time is added to the static model. Fig. 12 shows the dynamic model with three events.

Event 1 ($E_1$): The worker leaves for work.
Event 2 (E2): The worker goes to his or her regular duties.
Event 3 ($E_3$): The worker goes to work on the project.

The dynamic model shows the duration of the work time that is associated with Events 2 and 3. A "quantity" of time is, not surprisingly, related to time. Hence, the percentage of time is calculated from these thimacs of time, not from the relation between tasks. Fig. 13 shows the work-related behavior of the worker.

In general, events are viewed as "things that happen," which are mixed with other notions. As Worboys [34] stated, "One person's process is another's event, and vice versa." According to Galton [35],

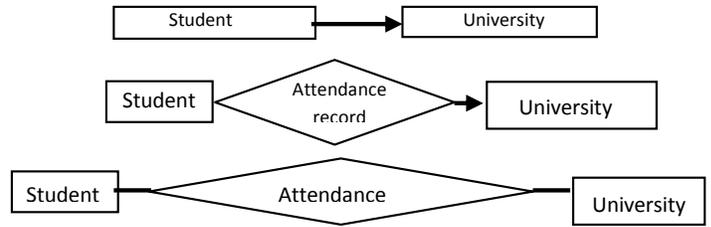

Fig. 9. Shorthand of the TM representation to produce the ER diagram.

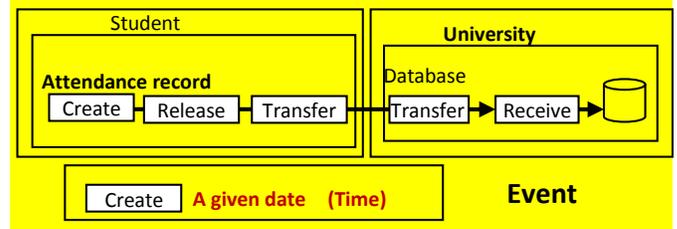

Fig. 10. The TM representation of the event *a student attends a university from a certain date*.

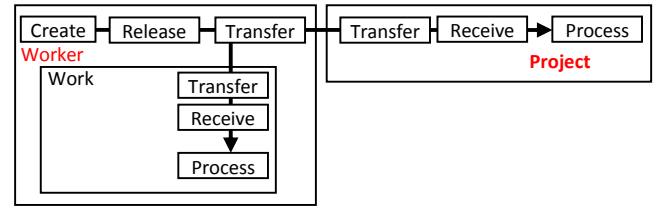

Fig. 11. The TM representation of *A worker who works on a project (in additional to his or her work*.

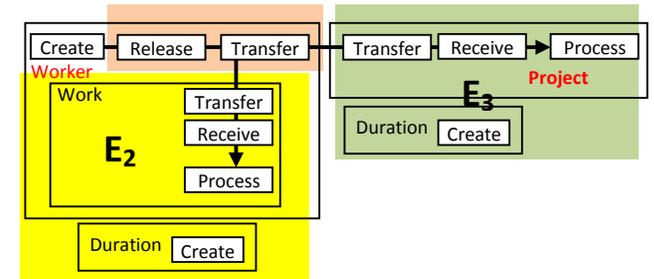

Fig. 12. The dynamic TM representation of *A worker who works on a project (in additional to his or her work)*.

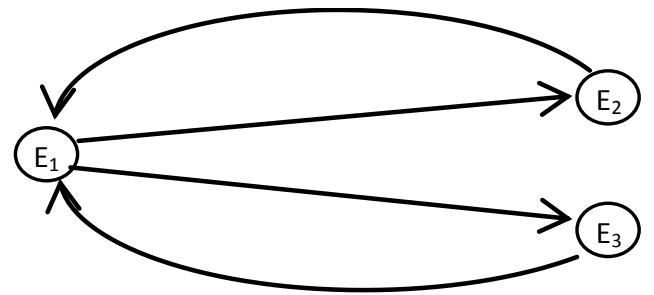

Fig. 13. The behavior of *a worker who works on a project*.

- An event is a "one-off" occurrence with a definite beginning and end; it has a completion and typically lasts a short time.
- A process is open-ended, continues indefinitely, and need never reach a state of completion.

In TM modeling, an event is "thimacs that happen," "a machine (process) that happens" (contains a time subthimac), and, simultaneously, "a thing that flows." A process is not necessarily an event. The sentence *He walks* may indicate that a person has the *capability* of walking (process), whereas *He walks this afternoon* describes an event (process plus time). In many modeling approaches, modelers mix up an operation/activity (e.g., walks) and an event. According to Guarino and Guizzardi [9], "verbs are typically the language proxy for events (including actions)"; however, *Approval of purchasing requisition* is an event if it happens in a certain time.

In this context, we can consider the argument "that a relationship is an object in itself" [10]. Guarino and Guizzardi [10] differentiate between relation and relationship: "Relations hold (that is, relational propositions are true) in virtue of the existence of a relationship; relationships are therefore truth makers of relations (more exactly, they are truth makers of relational propositions)."

From the TM perspective, all that we see is a thimac with two subthimacs: the worker and the project. (Without loss of generality, we ignore the sub-subthimac work in worker). Accordingly, we have three thimacs: the thimac and its two subthimacs, worker and project, as illustrated in Fig. 14. In Fig. 14, the overarching thimac embraces the worker and project subthimac as things and as machines. The thimac is a thing that includes the worker and project subthimacs. The semantics of this thing could be *physical worker and physical project* or *digital worker and digital project* (i.e., a tuple). One wonders, in this thimac representation, whether these two aspects of the thimac are what Guarino and Guizzardi [10] call a relationship (thing/project) and relation (machine). It is clear that in the TM model, the ER relationships dissipate in the flow that connects the worker to the project. Dissipation here refers to no explicit appearance even though a sense of relatedness remains between the worker and the project, but this sense comes from the flow between the two thimacs. The worker as a thing "goes into" the project (a machine). Note that the source of relationality in this picture is the machine part of the thimac. The sense of relatedness also appears in the chronology of events: A worker going to work at time $t1$ implies (doing their regular duties or working on a project at time $t2$ that follows $t1$).

Consider again the previously mentioned problem of relations, in which Aristotle required relations to serve to *relate* two (or more) things regardless of where they stand in relation to each other, such as Aristotle example mentioned previously, *x is ahead of y*. Thus, a difference exists between relative terms such as "father" and "son" and others such as "mover" and "moved" or "head" and "headed." Fatherhood is conceived of as if it were a kind of medium connecting a father with his son [16]. As mentioned previously, Avicenna (980-1037) argued that fatherhood is in the father alone and not in the son, and the same thing holds of sonship. Therefore, one must hold that two relations exist.

Fig. 15 shows the TM model of the fatherhood–sonhood thimac. Two things (persons) flow to the thimac (3) to be processed (4) to create two subthimacs: fatherhood (5 and 6) and sonhood (7 and 8) that flow to the father (9) and son (10), respectively.

V. MAINTAINING THE AIRPLANE

According to Sanati, Mehrizy, Welborn, and Minaie [36] (see also [37]), an important type of constraint on a business is timing-based or temporal constraints. As an introduction to temporal constraints, they review a problem that has been expressed in extended ER models.

Consider the following example: A mechanic must receive specific types of training related to maintaining airplanes. Many types of airplane-maintenance training programs are available, and, in turn, the types of training that a mechanic receives are used to determine the types of maintenance service that the mechanic can perform on an airplane. A specific maintenance service may require that a mechanic receive more than one type of training. Yet, a specific type of training may be useful in providing more than a single maintenance service [36]. This description is modeled in the ER model shown in Fig. 16.

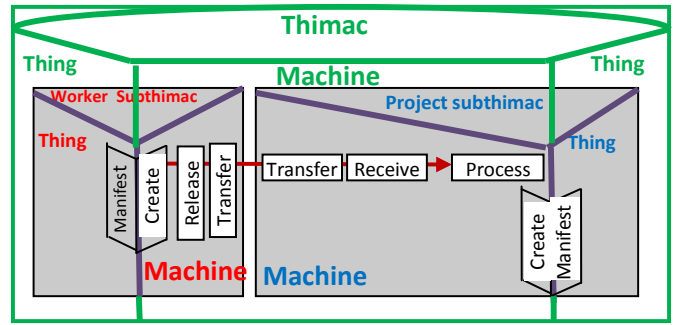

Fig. 14. The thimac that contains the worker and project subthimacs.

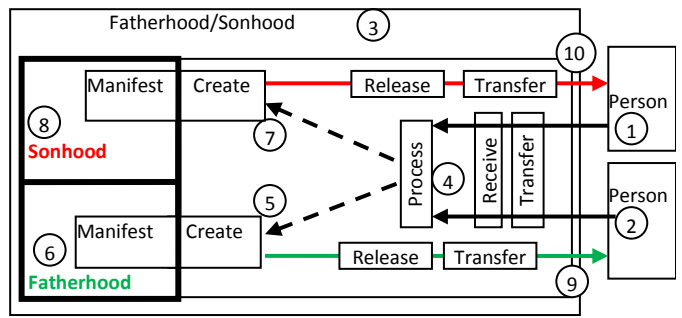

Fig. 15. The TM representation of the fatherhood/sonhood thimac.

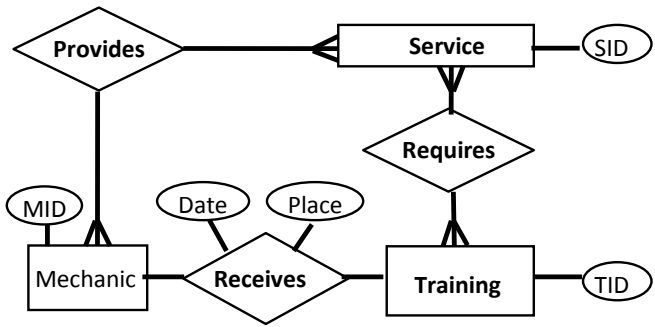

Fig. 16. The ER model of the airplane maintenance (adapted from [36]).

The business constraints of aircraft maintenance:

1. A mechanic can only provide a maintenance service if he or she has received all the training required for that service.
2. Maintenance service must be done using appropriate tools.
3. Maintenance service is only provided in a hanger.
4. A hangar can only allow service if two mechanics are present who can provide the same service [36].

Based on Sanati Mehrizy, Welborn, and Minaie [36], Fig. 16 does not guarantee that the mechanic will receive the required type of training for the type of service that he or she provides. Hence, they started a series of modified ER diagrams to handle such a problem. We will not pursue such modification and opt to model the example using the TM approach.

Figs. 17 shows the static TM model of this example. First, the airplane (1) moves to the hanger (2), where it is inspected (3). This triggers processing all services applicable to the airplane (4) to create a list of needed services (5). Needed services (6) and the type of trainings/services required (7) flow (8 and 9) where they are compared (10) to extract all trainings required for the needed services (11). All trainings required to maintain the airplane in addition to the mechanics' training (12; data of mechanics and their trainings) flow (13 and 14) to extract (15) a list of IDs of mechanics who have the required trainings (16). The processing of this list results in the selection of a team of two mechanics (17). The processing of this list (e.g., displaying the selected team) triggers the following:

- Forming the actual team (18; e.g., informing the mechanics)
- Releasing the airplane to the team (in the hanger) (19).
- Releasing the required services (20, 21, and 22).

The required services are processed (23) to check the availability of the tools for these services, and if the tools are available, then they are sent to the team of mechanics (24).

The dynamic model of Fig. 18 includes ten events identified as follows.

Event 1 ($E_1$): An airplane arrives in the hanger for maintenance.
Event 2 ($E_2$): Needed services are identified.
Event 3 ($E_3$): Corresponding trainings are identified.
Event 4 ($E_4$): Qualified mechanics are identified.
Event 5 ($E_5$): A two-mechanic team is selected.
Event 6 ($E_6$): The two-mechanic team is formed in the hanger.
Event 7 ($E_7$): The airplane is assigned to the team.
Event 8 ($E_8$): The team is informed of the needed repairs.
Event 9 ($E_9$): Required tools are identified.
Event 10 ($E_{10}$): The tools are sent to the team.

Fig. 10 shows the behavior of the maintenance system. The reflective arrows in the behavior indicate repetitive operations. For example, consider Event 2; assuming representation of tables, the event involves identifying a service. Assuming that a list of possible services exists, this event is repeated to select services to produce a list of needed services (see Fig. 20).

The TM representation is systematic in the sense that there is no need to invent a new notion to represent a new situation.

## VI. CONCLUSION

In this paper, several ER diagrams are re-casted as TM diagrams. The preliminary results indicate that the relationship is not necessary to a conceptual model since it is dissipated into TM flows of things and chronology of events. That is, the one TM ontology based on the thimac (thing+machine) suffices in producing a model of a portion of reality without the need for the notion of relation.

This seems to contradict statements in conceptual modeling that the relational construct is one of conceptual modeling's most fundamental constructs, as reviewed in the introduction. Similarly, such a conclusion seems to challenges Chen's [12] claim that "there are aspects of the intrinsic nature of 'the real-world' that would justify distinction between entities and relationship."

Of course, the conclusion here needs further scrutiny with more examples of ER diagrams remodeled using thimacs. The ER model may serve a different function than conceptual modeling at large (e.g., data modeling). Although the results seem to point in such a direction, we take the conclusion with caution until further experimentation of the involved type of modeling.

**References**

[1] S. G. Powell and K. R. Baker, Management Science: The Art of Modeling with Spreadsheets, 4th ed., Wiley, 2013.

[2] G. Scerra, "Problem Solving for Software Engineers," Code Project Site, Jan 1, 2015, https://www.codeproject.com/Articles/858726/Problem-Solving-for-Software-Engineers, accessed April 21, 2020.

[3] R. Kaschek, "On the evolution of conceptual modeling," Lecture Notes in Computer Science, 6520, State-of-the-Art Survey, LNCS Sublibrary, SL 3, Information Systems and Applications.

[4] J. Mylopoulos, "Conceptual modeling and Telos," in Conceptual Modeling, Databases, and CASE, ed. P. Loucopoulos & R. Zicari, pp. 49–68, Wiley, 1992.

[5] J. Mylopoulos, Philosophical Foundations of Conceptual Models, Centro de Informática, FUPE Recife, November 20, 2019. https://www.cin.ufpe.br/~in1020/docs/seminarios/John%20Mylopoulos%20-%20Seminar2.pdf, accessed Aril, 16, 2020

[6] A. Burton-Jones and R. Weber, "Understanding relationships with attributes in entity-relationship diagrams," ICIS 1999 Proceedings, vol. 20, 1999. https://aisel.aisnet.org/icis1999/20

[7] Y. Wand, V. C. Storey, and R. Weber, "An ontological analysis of the relationship construct in conceptual modeling," ACM Transactions on Database Systems, vol. 24, no. 4, pp. 494–528, December 1999.

[8] C. M. Fonseca, D. Porello, G. Guizzardi, J. P. A. Almeida, and N. Guarino, "Relations in ontology-driven conceptual modeling," International Conference on Conceptual Modeling, Salvador, Brazil, pp. 28-42, 2019.

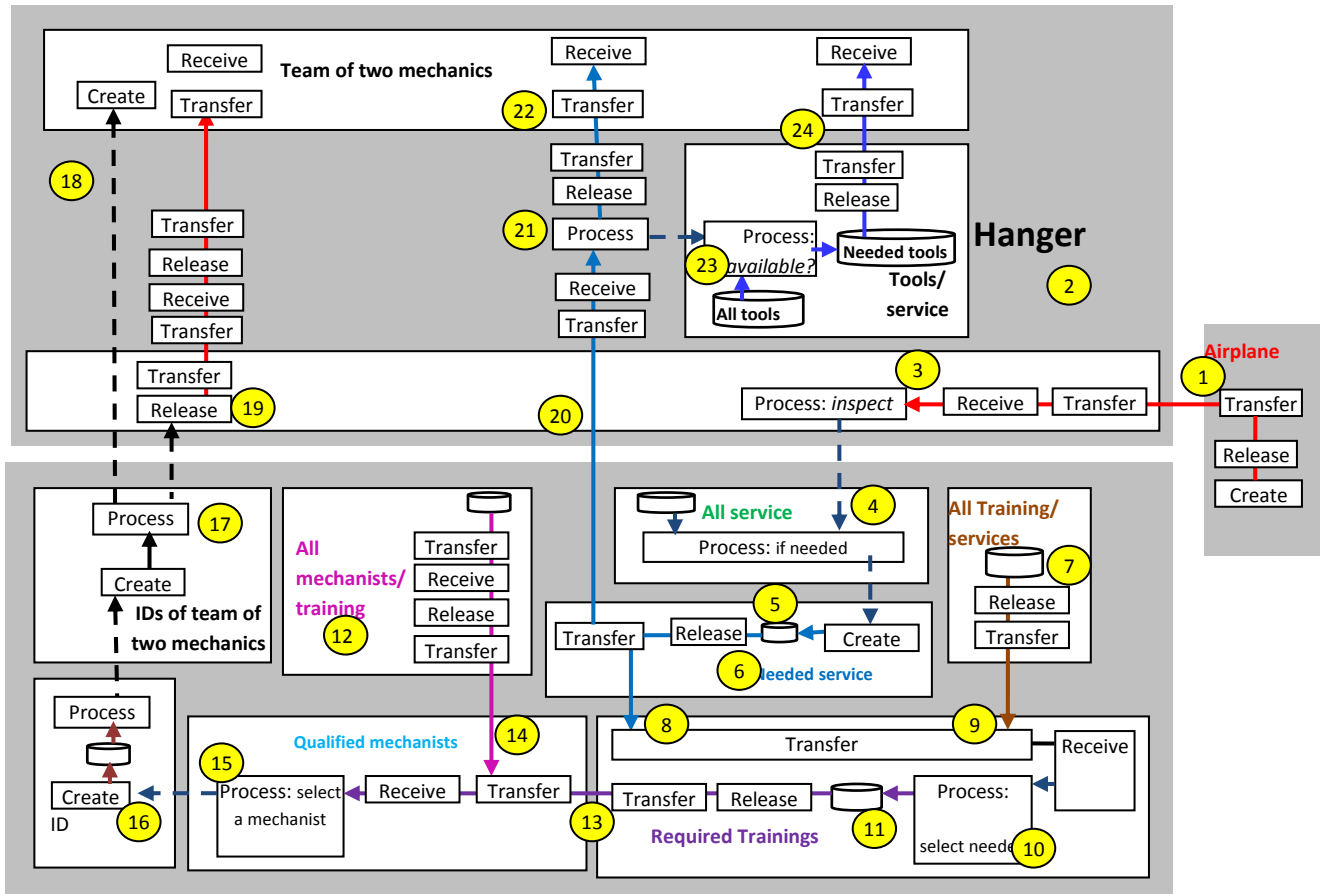

Fig. 17. The TM static model of the airplane maintenance example.

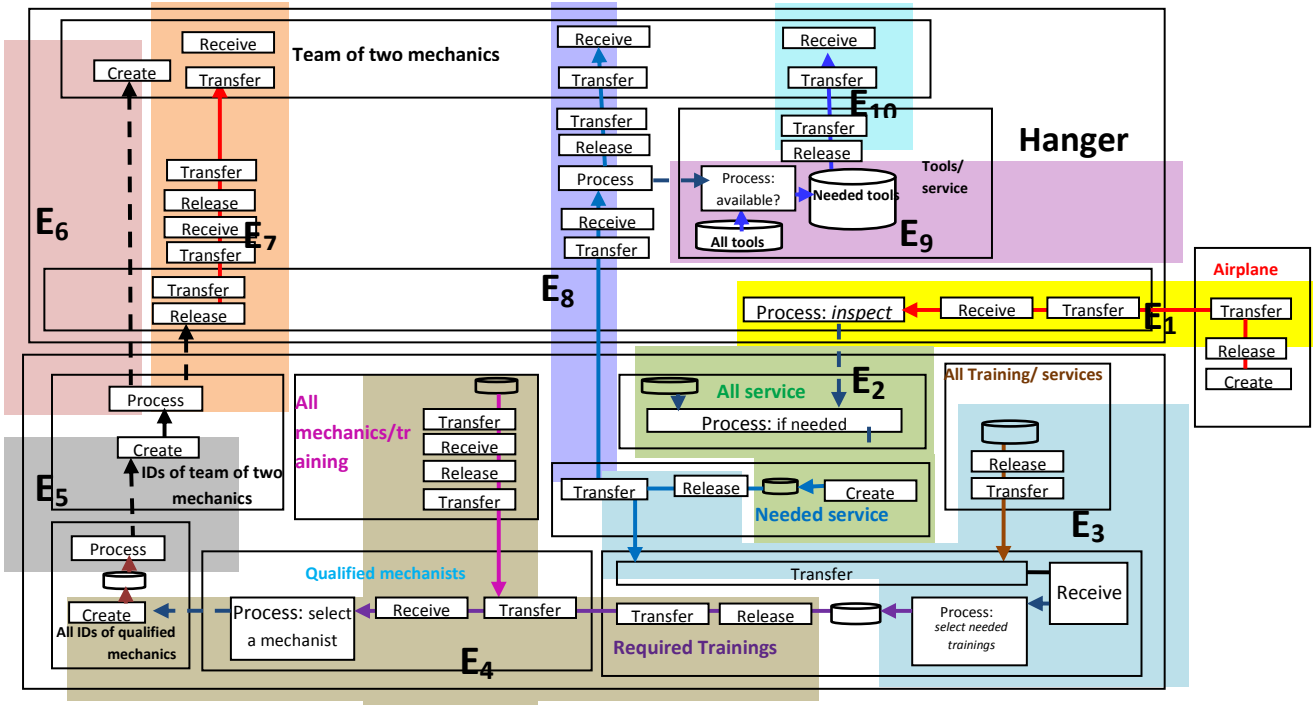

Fig. 18. The TM dynamic model of the airplane maintenance example.

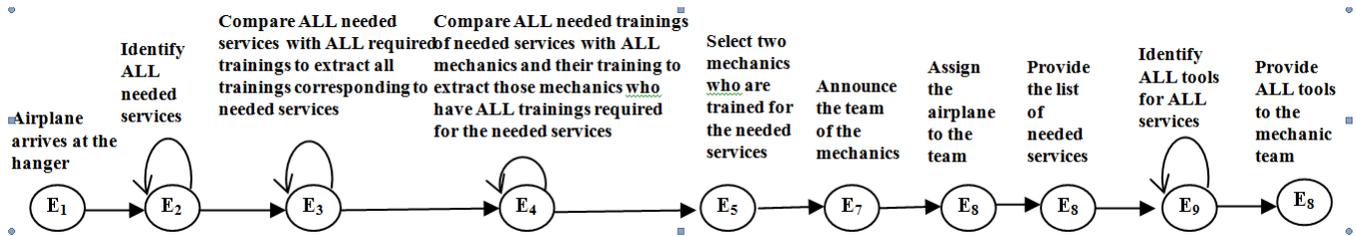

Fig. 19. The TM behavioral model of the airplane maintenance example.

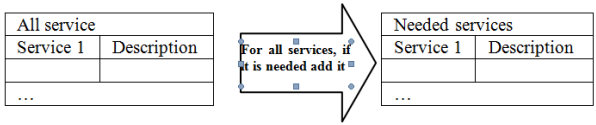

Fig. 20. Sample of a repeated event.

[9] N. Guarino and G. Guizzardi, G. (2015) "'We need to discuss the relationship': revisiting relationships as modeling constructs," in Advanced Information Systems Engineering, eds. J. Zdravkovic, M. Kirikova, and P. Johannesson, pp. 279-294, Spring, 2015. Lecture Notes in Computer Science, vol. 9097.

[10] N. Guarino and G. Guizzardi, "Relationships and Events: Towards a General Theory of Reification and Truthmaking," in AI*IA 2016 Advances in Artificial Intelligence, eds. G. Adorni, S. Cagnoni, M. Gori, and M. Maratea, pp. 237-249, Springer 2016. Lecture Notes in Computer Science, vol. 10037.

[11] W. J. Wildman, "An introduction to relational ontology," in The Trinity and an Entangled World: Relationality in Physical Science and Theologyn, eds. J Polkinghorne and J Zizioulas, pp. 55-73, Eerdmans, Grand Rapids, 2010.

[12] P. P. Chen, "The entity-relationship model: Toward a unified view of data." ACM Trans. Database Syst., vol. 1, no. 1, pp. 9–36, 1976.

[13] B. Thalheim, Entity-Relationship Modeling: Foundations of Database Technology, Springer-Verlag, Berlin, 2000.

[14] T. R. G. Green and D. R. Benyon, "The skull beneath the skin: entity-relationship models of information artefacts," International Journal of Human-Computer Studies, vol. 44, no. 6, pp. 801-828, 1996.

[15] D. Hay, "From data modeling to ontologies: discovering what exists, part 1," The Data Administration Newsletter, September 1, 2012, https://tdan.com/from-data-modeling-to-ontologies-discovering-what-exists-part-1/16331

[16] J. Brower, "Medieval theories of relations," in The Stanford Encyclopedia of Philosophy, ed. E. N. Zalta, Winter 2018, https://plato.stanford.edu/archives/win2018/entries/relations-medieval/, accessed April, 3, 2020.

[17] Al-Fedaghi, S. and Behbehani B., "How to document computer networks," Journal of Computer Science, Vol. 16. No. 6, pp. 723-434, 2020. DOI: 10.3844/jcssp.2020.723.434

[18] S. Al-Fedaghi, "Thinging Ethics for Software Engineers," International Journal of Computer Science and Information Security, Vol. 16, No. 9, September 2018.

[19] Al-Fedaghi, S. and Haidar E., 2020. "Thinging-based conceptual modeling: Case study of a tendering system," Journal of Computer Science, Vo. 16, No. 4, pp. 452-466. DOI: 10.3844/jcssp.2020.452.466

[20] S. Al-Fedaghi, "Modeling events and events of events in software engineering," International Journal of Computer Science and Information Security, vol. 18, no. 1, pp. 8-19, August 2019.

[21] S. Al-Fedaghi and D. Al-Qemlas, "Modeling network architecture: a cloud case study," International Journal of Computer Science and Network Security, vol.20, no.3, pp.195-209, March 2020.

[22] S. Al-Fedaghi, "Modeling Physical/Digital Systems: Formal Event-B vs. Diagrammatic Thinging Machine," International Journal of Computer Science and Network Security, vol.20, no.4, pp.208-220, April 2020.

[23] S. Al-Fedaghi, "Modeling communication: one more piece falling into place," The 26th ACM International Conference on Design of Communication (SIGDOC 2008), Lisbon, Portugal, Sept. 22-24, 2008.

[24] P. van Inwagen, "Relational vs. constituent ontologies, philosophical perspectives," Metaphysics, vol. 25, 389-402, 2011.

[25] L. A. Paul, A One Category Ontology, In Being, Freedom, and Method: Themes from the Philosophy of Peter van Inwagen, ohn A. Keller, Oxford Scholarship Online, Oxford University Press UK, January 2017.

[26] H. G. Steiner, "Theory of mathematics education: An introduction," For the Learning of Math., vol. 5, no. 2, pp. 11–17, 1985.

[27] A. Sfard, "On the dual nature of mathematical conceptions: Reflections on processes and objects as different sides of the same coin," Educ. Studies in Math., vol. 22, no. 1, pp. 1–36, 1991.

[28] J. Mason and A. Waywood, "The role of theory in mathematics: education and research," in International Handbook of Mathematics Education, eds. A. Bishop, M.A. Clements, C. Keitel-Kreidt, J. Kilpatrick, and C. Laborde, pp. 1055-1089, Springer Science and Business Media, December 6, 2012.

[29] M. Heidegger, "The thing," in Poetry, Language, Thought, trans. A. Hofstadter, pp. 161–184, Harper and Row, New York, 1975.

[30] S. Ash, Funding Philanthropy: Dr. Barnardo, Metaphor, Narrative and Spectacle, Oxford University Press, Jun 1, 2016.

[31] L. K. Rhodes, Lecture notes, Juniata College, Huntingdon, PA. http://jcsites.juniata.edu/faculty/rhodes/dbms/ermapping.htm, accessed Accessed May, 7, 2020.

[32] Y. Wand and R. Weber, "An ontological analysis of the relationship construct in conceptual modeling," ACM Transactions on Database Systems, vol. 24, no. 4, pp. 494–528, December 1999.

[33] F. P. Fuhs, "Event-extended entity-relationship diagrams for understanding simulation model structure and function." Dev. Bus. Simul. Exp. Exerc., vol. 15, pp. 41–45, 1988.

[34] M. F. Worboys, "Event-oriented approaches to geographic phenomena," International Journal of Geographical Information Science, vol. 19, no. 1, pp. 1–28, 2005.

[35] A. Galton, "The ontology of time and process," Third Interdisciplinary School on Applied Ontology, Bozen-Bolzano, June 27–July 1, 2016.

[36] R. Sanati Mehrizy, C. Welborn, and A. Minaie, "Representing and enforcing business rules in relational data model," American Society for Engineering Education, 2006.

[37] C. Welborn and R. Sanati-Mehrizy, "Temporal extensions for enhanced entity relationship notation," American Society for Engineering Education Annual Conference and Exposition, Pittsburgh, Pennsylvania, June 22-25, 2008.